\documentclass[preprint]{revtex4-1}
\usepackage{graphicx} 
\usepackage{amsmath}
\usepackage{caption}
\usepackage{float}
\usepackage{mathptmx}
\usepackage{comment}
\usepackage{natbib}
\usepackage{enumitem}
\usepackage[colorlinks]{hyperref}
\usepackage{tabularx}
\usepackage{dcolumn}
\usepackage{bm}
\begin{document}
\preprint{APS/Physical Review Fluids}

\title{Turbulence closure modeling with data-driven techniques:\\physical compatibility and consistency considerations}
\author{Salar Taghizadeh\textsuperscript{1}}
\email{staghizadeh@tamu.edu}
\author{Freddie D. Witherden\textsuperscript{2}}
\author{Sharath S. Girimaji\textsuperscript{1,2,3}}
\affiliation{\footnotesize $^1$Department of Mechanical Engineering, Texas A\&M University, College Station, TX 77843\looseness=-1}
\affiliation{$^2$Department of Ocean Engineering, Texas A\&M University, College Station, TX 77843}
\affiliation{$^3$Department of Aerospace Engineering, Texas A\&M University, College Station, TX 77843}

\begin{abstract}
A recent thrust in turbulence closure modeling research is to incorporate machine learning (ML) elements, such as neural networks, for the purpose of enhancing the predictive capability to a broader class of flows. Such a turbulence closure framework entails  solving a system of equations comprised of ML functionals coupled with traditional (physics-based - PB) elements. While combining closure elements from fundamentally different ideologies can lead to unprecedented progress, there are many critical challenges that must be overcome. This study examines three such challenges: (i) Physical compatibility (or lack thereof) between ML and PB constituents of the modeling system of equations; (ii) Internal (self) consistency of the ML training process; and (iii) Formulation of an optimal objective (or loss) function for training. These issues are critically important for generalization of the ML-enhanced methods to predictive computations of complex engineering flows. Training and implementation strategies in current practice that may lead to significant incompatibilities and inconsistencies are identified. Using the simple test case of turbulent channel flow, key deficiencies are highlighted and proposals for mitigating them are investigated. Compatibility constraints are evaluated and it is demonstrated that an iterative training procedure can help ensure certain degree of consistency. In summary, this work develops  foundational tenets to guide development  of ML-enhanced turbulence closure models.

\end{abstract}
\begin{titlepage}
\maketitle
\end{titlepage}                      
\section{\label{S:1}Introduction}
Many studies in recent literature aim to enhance the capabilities of turbulence models by incorporating data-driven techniques.  These methods seek to better utilize the increasingly available high fidelity numerical data in turbulent flows along with recent developments in machine learning (ML) to improve turbulence model predictions in unseen complex engineering flows. In these approaches, functionals obtained using ML are used within existing turbulence modeling framework to incorporate capabilities that fall beyond the scope of traditional approaches. In principle data-driven methods may be used to improve model performance at any level of turbulence closure including large-eddy simulations (LES), scale-resolving simulations (SRS) and Reynolds-averaged Navier--Stokes (RANS) methods. 

In their seminal work, Sarghini et al. \cite{sarghini2003neural} adopted neural networks 
to develop improved subgrid scale models for LES of turbulence. 
Gamahara and Hattori \cite{gamahara2017searching}, Maulik and San \cite{maulik2017neural}, 
Maulik et al. \cite{maulik2018data} and Beck et al. \cite{beck2019deep} proposed ML-enhanced LES closures using neural networks. Different ML techniques are also adopted to enhance RANS turbulence models. Random forest technique \cite{breiman2001random} has been used to improve RANS turbulence models by Wang et al. \cite{wang2017physics} and 
Wu et al. \cite{wu2017priori}. Field inversion and neural networks techniques are employed to introduce correction factors in RANS modeled transport equations by Singh and Duraisamy \cite{singh2016using}, Singh et al. \cite{singh2017machine}, 
Zhang et al. \cite{zhang2015machine} and Parish and Duraisamy \cite{parish2016paradigm}. 
Galilean invariant Reynolds stress anisotropy models are trained using tensor basis neural networks by 
Ling et al. \cite{ling2016reynolds}. Explicit algebraic Reynolds stress models have been developed using gene expression 
programming (GEP) by Weatheritt and Sandberg \cite{weatheritt2016novel} and applied to different test flows by 
Weatheritt and Sandberg \cite{weatheritt2017development}, Weatheritt et al. \cite{weatheritt2017machine}, 
Akolekar et al. \cite{akolekar2018development} and Zhao et al. \cite{zhao2019turbulence}. 
GEP is also applied in unsteady RANS and PANS (partially averaged Navier-Stokes)  simulations 
by Lav et al. \cite{lav2019framework}. 
A review of the important studies in this area can be found in Kutz \cite{kutz2017deep} and 
Duraraisamy et al. \cite{duraisamy2019turbulence}.

Incorporating data-driven techniques into the turbulence closure modeling process can have a transformative influence on the field. While preliminary studies show basis for optimism, more research is needed to understand the physical underpinnings of 
data-driven methods in order to ensure their generalizability to unseen test flows.
In order to maximize their impact, data-driven approaches must leverage the physical understanding and closure modeling knowledge already incumbent in traditional models. 
Two-equation RANS models are still the most widely used tools in practical flow calculations and any improvement of their capabilities can have significant impact on engineering applications. Towards this end, the goal of this work is to investigate data-driven approaches for RANS enhancement.
 For the sake of clarity and contrast,  through the remainder of this paper, we denote traditional closures with the prefix PB (to indicate physics-based) and the novel data-driven approach with  ML (for machine learning). 
 
PB-RANS computations of a turbulent flow involves the solution of a dynamically interacting system of equations. The two-equation RANS model is often called the lowest-order complete closure model \cite{wilcox1998turbulence} as it solves independent model equations for length and velocity scales to compute eddy viscosity. There are three main closure elements in a two-equation RANS model: algebraic (linear or non-linear) Reynolds stress constitutive relation; a modeled transport (partial-differential) equation for kinetic energy to provide the turbulence velocity scale; and, a modeled transport (partial-differential) equation for dissipation or turbulence frequency to specify the turbulence length scale. The closure models and coefficients are typically developed in canonical flows that highlight individual turbulence processes. For reliable predictive computations of complex flows, the individual models must be independently accurate, and even more importantly, the dynamical interplay between the various equations must be compatible and consistent with overall flow physics. In PB-methods, the required compatibility between the various equations is accomplished (to the extent possible) by performing a dynamical system analysis. The fixed-point behavior at various asymptotic limits is examined for consistency with known physics. Such a systematic strategy assures some degree of generalizability to unseen flows. 

Data-driven approaches for two-equation RANS proposed in literature employ machine learning (ML) methods for certain closures and retain traditional models for other aspects. For example Ling et al. \cite{ling2016reynolds} and Weatheritt and Sandberg \cite{weatheritt2016novel, weatheritt2017development} use ML for obtaining improved Reynolds stress constitutive relations while the modeled transport equations for turbulence length and velocity scales are retained without changes. On the other hand, Zhang et al. \cite{zhang2015machine} and Parish and Duraisamy \cite{parish2016paradigm} use ML to optimize transport equation coefficients for best performance in flows of their interest without changing Reynolds stress constitutive relation. Thus the data-driven closure framework represents a mix of ML and traditional (PB) models. For generalizability of these methods to untested flows, it is important that the data-driven framework satisfy key requirements. In this study, we examine three such prerequisites:
\begin{enumerate}
 \item \textbf{\emph{Physical Compatibility:}} In the PB-RANS model the various coefficients are carefully orchestrated to yield reasonable and holistic behavior in a set of canonical cases. Any ML-based modification of a subset of these coefficients can have deleterious effect on the overall computed outcome. Therefore, the importance of ensuring compatibility between ML functionals and PB elements is investigated.
 \item \textbf{\emph{Training  Consistency:}} ML training requires input features that are currently obtained from baseline RANS models which employ PB-closure coefficients. Then, the training produces an improved functional for some of the same closure coefficients. In current methods, there is no process to ensure consistency between the {\em a priori} PB-closure coefficients that produce the input features and the {\em a posteriori} ML values. The consequences of such inconsistency is examined and means of imposing consistency are proposed.
 \item \textbf{\emph{Loss function formulation:}} The success of the ML training approach hinges on the formulation of an appropriate loss function. The challenges in identifying the optimal flow statistics contributing to the loss function are examined.
\end{enumerate}
It bears repeating that the goal of the study is not to propose a specific ML closure approach, but it is to establish fundamental guidelines for ML-RANS model development. 

The paper is organized as follows. The RANS closure framework is discussed in Sec.~\ref{sec:RANS}. Key challenges in applying ML techniques to two-equation turbulence model are identified in Sec.~\ref{sec:ML}. A closed loop training framework is proposed in this section. Section \ref{sec:Proof} formulates proof-of-concept studies and demonstrates importance of defining appropriate loss function for ML. The results and inferences are presented in Sec.~\ref{sec:Find}. Finally, the conclusions of this study are summarized in Sec.~\ref{sec:Conclusion}.

\section{\label{sec:RANS}RANS Closure framework}

The Navier--Stokes equations for a viscous and incompressible flow can be written as, 
\begin{equation}  
\frac{\partial V_i}{\partial x_i}=0, \quad
\frac{\partial{V_i}}{\partial t}+V_j\frac{\partial V_i}{\partial x_j}=-\frac{1}{\rho}\frac{\partial p}{\partial x_i}+\nu \frac{\partial^2 V_i}{\partial x_j \partial x_j},
\label{eq:NS}
\end{equation}
where $V_i$ is the instantaneous velocity, $p$ is the instantaneous pressure, $\rho$ is the density and $\nu$ is the kinematic viscosity. Upon applying the Reynolds averaging operator \cite{reynolds1895iv} on the Navier--Stokes equations, RANS equations for incompressible flows are obtained:
\begin{equation}  
\frac{\partial U_i}{\partial x_i}=0,\quad 
\frac{\partial{U_i}}{\partial t}+U_j\frac{\partial U_i}{\partial x_j}=-\frac{1}{\rho}\frac{\partial P}{\partial x_i}-\frac{\partial \langle u_iu_j \rangle }{\partial x_j}+\nu \frac{\partial^2 U_i}{\partial x_j \partial x_j},
\label{eq:NS_Avg}
\end{equation}
where $U_i$ is the mean velocity and $P$ is the mean pressure. The Reynolds stress tensor $(\langle u_iu_j\rangle)$ in this equation is the subject of closure modeling. This symmetric, second order tensor can be decomposed into isotropic and anisotropic $(b_{ij})$ parts, 
\begin{equation}  
\langle u_iu_j \rangle=-\tau_{ij}=\frac{2}{3}\delta_{ij}k + 2kb_{ij}, \quad k=\frac{1}{2}\langle u_ku_k \rangle.
\label{eq:RS}
\end{equation}
where $k$ is the turbulent kinetic energy, and $\delta _{ij}$ is the Kronecker delta. In Reynolds stress closure modeling (RSCM) approach, modeled transport equations are solved for all independent components of Reynolds stress tensor $(\langle u_iu_j \rangle)$ \cite{launder1975progress, speziale1991modelling, johansson1994modelling, girimaji2000pressure}. At the current time, ML methods have not been employed  much for enhancing RSCM closure approach. 

\subsection{\label{sec:two-eqn}Two-equation RANS}
In the two-equation RANS approach, which is the subject of this study, a constitutive relationship for Reynolds stress is postulated in terms of the strain and rotation rates of the mean flow field. Modeled transport equations are solved for turbulence velocity and length scales to yield eddy viscosity.

\textbf{Reynolds stress constitutive relationship:} Using representation theory, a general form of the constitutive relationship for the normalized anisotropic tensor can be written as \cite{pope1975more},

\begin{equation}  
b_{ij}(s_{ij}, r_{ij})=\sum_{n=1}^{10} G_n (\lambda_1,..., \lambda_5, k, \epsilon) T^{(n)}_{ij},
\label{eq:Represent.}
\end{equation}
in terms of ten basis tensors $T^{(n)}_{ij}$ and their scalar invariant functions $\lambda_1,..., \lambda_5$. Here $\epsilon$ is the turbulent dissipation. The basis tensors and scalar invariants are known functions of the normalized mean strain $(s_{ij})$ and rotation rates $(r_{ij})$ \cite{pope1975more},
\begin{equation}  
  s_{ij}=\frac{k}{\epsilon}S_{ij},\quad
  r_{ij}=\frac{k}{\epsilon}R_{ij}.
\label{eq:Strain}
\end{equation}
where 
\begin{equation}  
  S_{ij}=\frac{1}{2}(\frac{\partial U_i}{\partial x_j}+\frac{\partial U_j}{\partial x_i}),\quad
  R_{ij}=\frac{1}{2}(\frac{\partial U_i}{\partial x_j}-\frac{\partial U_j}{\partial x_i}).
\label{eq:Strain2}
\end{equation}
The scalar function $G_{n}$ of each basis tensor $T^{(n)}_{ij}$ must be modeled. Through the reminder of this study the \emph{G's} are referred to as constitutive closure coefficients (CCC). Different Reynolds stress constitutive relations of varying degree of complexity have been proposed in literature. The simplest constitutive relation is the Boussinesq model \cite{wilcox1998turbulence} given by, 
\begin{equation}  
b_{ij} = -C_{\mu} s_{ij}, 
\label{eq:bij}
\end{equation}  
and turbulent viscosity ($\nu_t$) can be written as, 
 \begin{equation}
    \nu_t=C_{\mu}\frac{k^2}{\epsilon}.
\label{eq:nut}
\end{equation}
In this model, the CCC are: $G_1=-C_{\mu}= -0.09$ and $G_n=0$ for $n>1$ \cite{pope2001turbulent}. More complex non-linear eddy viscosity models \cite{speziale1987nonlinear, craft1996development} and algebraic Reynolds stress models (ARSM) \cite{pope1975more, rodi1976new, gatski1993explicit, girimaji1996fully} have also been proposed. Various non-linear and ARSM models determine the \emph{G's} using different approaches to match equilibrium anisotropies in various flows.

The goal of ML-enhancement is to learn ML functionals for CCC using  high fidelity data in flows of choice. 

\textbf{Modeled transport equations:} The required turbulence velocity and length scales are obtained by solving the modeled transport equations for turbulent kinetic energy $(k)$ and dissipation $(\epsilon)$ or specific dissipation rate $(\omega=\frac{\epsilon}{{\beta}^*k})$. The standard $k-\omega$ modeled transport equations are:
\begin{equation}  
\begin{split}
\frac{\partial k}{\partial t} + U_j\frac{\partial k}{\partial x_j} &= \tau_{ij}\frac{\partial U_i}{\partial x_j}-\beta^*k\omega + \frac {\partial} {\partial x_j} \Big[(\nu+\sigma^*\nu_t)\frac{\partial k}{\partial x_j}\Big],\\
\frac{\partial \omega}{\partial t} + U_j\frac{\partial \omega}{\partial x_j} &= \alpha \frac{\omega}{k}\tau_{ij}\frac{\partial U_i}{\partial x_j}-\beta \omega^2 + \frac {\partial} {\partial x_j} \Big[(\nu+\sigma \nu_t)\frac{\partial \omega}{\partial x_j}\Big].
\end{split}
\label{eq:ktransport}
\end{equation}
Here $\alpha$, $\beta$, $\beta^*$, $\sigma$ and $\sigma^*$ are the transport closure coefficients (TCC). In traditional modeling, the values of the TCC are determined to satisfy known asymptotic or equilibrium behavior in canonical flows. Each calibration flow is chosen to highlight a key turbulence process.

\textbf{\emph{Decaying Isotropic Turbulence (DIT):}} 
Decaying homogeneous isotropic turbulence is the simplest non-trivial turbulent flow wherein production and transport terms vanish and there is no spatial variation of the flow statistics. This case is used to determine the ratio $\frac{\beta}{\beta^*}$ from the decay rate of turbulent kinetic energy \cite{wilcox1998turbulence}. The modeled transport equations for $k$ and $\omega$ (Eq.~\eqref{eq:ktransport}) reduces to:
\begin{equation}  
\frac{\partial k}{\partial t} = -\beta^*k\omega,\quad \frac{\partial \omega}{\partial t} = -\beta \omega^2,
\label{eq:DIC}
\end{equation}
leading to the following asymptotic power-law decay of kinetic energy and turbulence frequency:
\begin{equation}
    k(t) = k_0\Big(\frac{t}{t_0}\Big)^{-n},  \quad \omega(t) 
    =\omega_0\Big(\frac{t}{t_0}\Big)^{-1},\;\; \mbox{ where } \;\; n = \frac{\beta^*}{\beta}.
\label{eq:Asymtote}
\end{equation}
In the above equation $k_0$, $\epsilon_0$ and $\omega_0$ are values for $k$, $\epsilon$ and $\omega$ at the reference time $t_0=n\frac{k_0}{\epsilon_0}$, respectively. 
It is known from a variety of experiments and DNS that the kinetic energy power-law decay exponent \emph{n} is nearly a constant in the range $1.15<n< 1.45$. In standard $k-\omega$ model, the ratio $\frac{\beta}{\beta^*}$ is determined by selecting \emph{n} = 1.25,
\begin{equation}  
\frac{\beta}{\beta^*} = \frac{1}{1.25}=0.8.\\
\label{eq:betaN}
\end{equation}

\textbf{\emph{Equilibrium behavior of homogeneous turbulence:}} 
In a homogeneous flow, production term is non-zero and all transport terms are still zero. In  energetic homogeneous turbulent flows three key dimensionless quantities --  turbulence frequency ($\omega$), production-to-dissipation ratio ($P/\epsilon$) and mean-to-turbulence frequency ratio ($Sk/\epsilon$)--  evolve to their respective equilibrium states. The equilibrium values of the 
these quantities can be related to the unclosed model coefficients (TCC) by performing a fixed point analysis of Eq.~\eqref{eq:ktransport}:
\begin{equation}  
\frac{\partial \omega}{\partial t} =\alpha \frac{\omega}{k}P-\beta \omega^2 = 0.
\label{eq:Eq.omega}
\end{equation}
Invoking the definition of the turbulence dissipation $(\epsilon={\beta}^*k\omega)$, Eq.~\eqref{eq:Eq.omega} can be simplified to yield, 
\begin{equation}  
\frac{P}{\epsilon} = \frac{\beta }{\alpha \beta^*}.
\label{eq:Omega-Producion}
\end{equation}
Employing Eq.~\eqref{eq:nut}, the production term \emph{(P)} can be written as, 
\begin{equation}  
P = \tau_{ij}\frac{\partial U_i}{\partial x_j} = \nu_t S^2 = \frac{C_\mu k^2}{\epsilon} S^2 = \frac{-G_1 k^2}{\epsilon} S^2,
\label{eq:Producion}
\end{equation}
whers \emph{S} is defined as $S \equiv \sqrt{2S_{ij} S_{ij}}$. Equation \eqref{eq:Producion} can be rewritten as, 
\begin{equation}  
\frac{P}{\epsilon} = -G_1\Big(\frac{Sk}{\epsilon}\Big)^2.
\label{eq:ProducionDis}
\end{equation}
Using Eqs.~\eqref{eq:Omega-Producion} and \eqref{eq:ProducionDis}, the following  relationship can be obtained, 
\begin{equation}  
\Big(\frac{Sk}{\epsilon}\Big)^2 = \frac{\beta}{-G_1\alpha \beta^*}.
\label{eq:ProductionDissipation}
\end{equation}
The fixed-point solutions relate $\alpha$ and $\beta$ to the equilibrium values of $Sk/\epsilon$ and $P/\epsilon$. From a suite of experiments and numerical simulations of homogeneous turbulence, it is known that the range of mean-to-turbulence frequency ratio is $4.0<Sk/\epsilon< 6.5$ \cite{pope2001turbulent}. The production-to-dissipation ratio range is $1.5<P/\epsilon< 2.0$ \cite{pope2001turbulent}. 
In standard $k-\omega$ model, $Sk/\epsilon=4.13$ and $P/\epsilon=1.54$ are selected to determine the values for $\alpha$ and  $\beta$ coefficients using Eqs.~\eqref{eq:Omega-Producion} and \eqref{eq:ProductionDissipation}.

Body forces and other factors such as reference frame rotation and streamline curvature can change the above equilibrium behavior of turbulence. For turbulence in a rotating frame, Speziale et al. \cite{speziale1998consistency}, demonstrate that dissipation and eddy viscosity (CCC values) must vanish asymptotically in the limit of infinite rotation. Thus there are many implied or explicit relationships between TCC and CCC. The exact nature of the relationship between TCC and CCC will depend on the order of the constitutive relation (e.g., linear, quadratic, cubic) and the complexity of flow (e.g., rotating reference frame) and type of flow.

\textbf{\emph{Equilibrium behavior of log-layer:}} Turbulent transport model coefficients ($\sigma$ and $\sigma^*$) are developed from the analysis of equilibrium boundary layer \cite{wilcox1998turbulence}:
\begin{equation}
    \sigma=\frac{\sqrt{-G_1}(\frac{\beta}{\beta^*}-\alpha)}{\kappa^2}.
\label{eq:sigma}
\end{equation}
In the above  $\kappa$ is the Karman constant.  Literature suggests that the value of this constant is in the range $0.384<\kappa< 0.41$ \cite{chauhan2007evidence}. 

In the equilibrium log-layer, the Reynolds shear stress and kinetic energy are related as follows: \cite{wilcox1998turbulence}, 
\begin{equation}
    \tau_{xy}=\sqrt{\beta^*}\rho k.
\label{eq:betaS}
\end{equation}
It has been shown  $\beta^*=0.09$ leads to a log-layer solution consistent with experimental measurements \cite{wilcox1998turbulence}.

Based on the above analyses, the TCC for the standard $k$-$\omega$ model (assuming $G_1 = - 0.09)$ are
specified as: 
\begin{equation}
   \alpha = 0.52;  \quad \beta = 0.072; \quad \beta^* = 0.09; \quad \sigma = \sigma^* = 0.5.
\end{equation}
In the above, the Karman constant is taken to be $0.41$ \cite{pope2001turbulent}.
It is evident from the above discussion that the transport equation closure coefficients (TCC) and the  the Reynolds stress constitutive equation coefficients (CCC) are strongly interconnected. Any change in a subset of coefficients without corresponding modification of others can lead to erroneous model behavior. It is essential that any closure procedure must make allowance for these relationships in the model development process.  
 
While the two-equation models with advanced Reynolds stress constitutive relations perform adequate in some applications of engineering interest, the models are generally found wanting in flows which include complexities not accounted for in the model derivation procedure, e.g., non-equilibrium turbulence states, largescale unsteadiness, underlying instabilities, and spatially-developing features.

\section{\label{sec:ML}Data Driven frameworks}
To improve the performance of two-equation RANS models in complex flows, several recent studies have considered replacing some traditional model elements by trained functionals obtained from ML. Proposed modifications include improving the Reynolds stress constitutive relations as described in \cite{ling2016reynolds, weatheritt2016novel, weatheritt2017development, weatheritt2017machine, akolekar2018development, lav2019framework, zhao2019turbulence}, and optimizing coefficients in the modeled transport equations as given in \cite{duraisamy2015new, tracey2013application, zhang2015machine, parish2016paradigm}. In both instances the resulting system of equations are composed of traditional closure elements and data-derived functionals obtains from ML. 
In what follows we describe the current training methodology which is the open loop approach. Then, we identify shortcomings and propose potential improvements.

\subsection{\label{sec:Open}Open loop framework}
\begin{figure}
    \centering
    \includegraphics[width=0.9\textwidth]{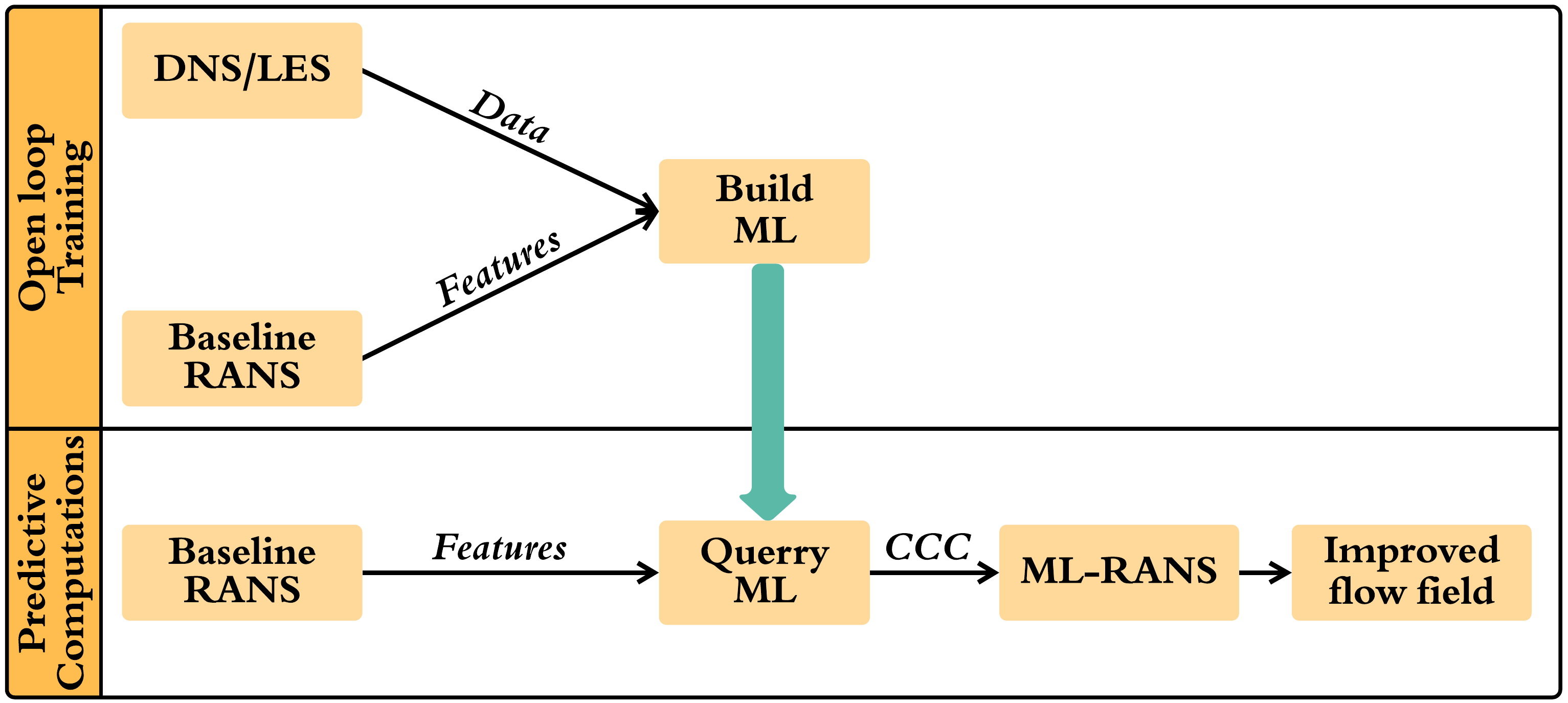}
    \caption{ Open loop framework}
    \label{Open}
\end{figure}

The ML approaches in many current studies employ \emph{open loop} training framework. Here we briefly describe the approach adopted by Ling et al. \cite{ling2016reynolds}. A schematic of the approach comprising of training and predictive computation stages is shown in Fig.~\ref{Open}. For {\bf {\it training (building)}} the ML functional, high fidelity data of the Reynolds stress tensor $(\langle u_iu_j\rangle)$ or the normalized anisotropy tensor $(b_{ij})$ is obtained from direct numerical simulations (DNS) or LES  of chosen flows. The input features for training ML functional are derived from the baseline (traditional) RANS model calculations. The features include contributions from the solutions of the mean flow equations ($s_{ij}$ and $w_{ij}$) along with those of $k$ and $\omega$ modeled transport equations. The ML functional training is then undertaken to identify the `best possible' Reynolds stress constitutive relation (optimal values of CCC) by minimizing a suitably defined loss function simultaneously over all the training cases. This leads to values of CCC that are modified from those of the baseline RANS case. However, the closure coefficients of the modeled transport equations (TCC) are retained at the original values.

{\bf{\it Predictive computations }} of previously unseen flow cases involve the following steps. In the first step, baseline RANS is once again performed to generate the features - mean flow strain rate, rotation rate, turbulent kinetic energy and dissipation fields. Then the ML functional is invoked to yield CCC values corresponding to the input features from the baseline RANS. 
The resulting `ML constitutive relation' is used within the RANS framework (in momentum equation and turbulent kinetic energy production term) to compute an updated flow field. It is anticipated that updated flow field will represent an improvement over the baseline RANS computation.
This ML-RANS procedure can be considered {\em open loop} as there is no feedback from the ML output to the training process. 

The current open loop procedure has two potential shortcomings, especially if the ML modification to the CCC is large:
\begin{enumerate}
  \item {\bf Physical incompatibility}:  The ML-trained values of CCC may not be compatible with the baseline values of TCC - see for example Eqs.~\eqref{eq:ProductionDissipation} and \eqref{eq:sigma}. This incompatibility can possibly lead to unphysical behavior of the model, even if the flow variables in the loss function behave reasonably well. To improve compatibility, one or both of the Eqs.~\eqref{eq:ProductionDissipation} and \eqref{eq:sigma} can be used to modify TCC values to be in accordance with ML values of CCC. However, over-constraining the coefficients may lead to other drawbacks. This will be discussed in the later sections.
  \item {\bf Training inconsistency}: The baseline model (with standard CCC) is used to generate input features for ML training.  The training process produces a functional for CCC which is used for predictive computations. The CCC values used to generate the input features for training will not be the same as those obtained from the trained ML functional.
  Thus there is inherent inconsistency between {\em a priori} and {\em a posterior} values of CCC   which can result in change in the dynamical character of the system of equations. As a result the fixed point behavior of quantities not included in the loss function can be significantly different. 
  The inconsistency is exacerbated if TCC values are also changed to satisfy  compatibility constraints. To improve consistency between {\em a priori} and {\em a posteriori} -- before and after ML training -- values of the coefficients, we propose a closed loop training method. 
\end{enumerate}

In summation, the physical inconsistency and incompatibility of the current open loop training procedure can adversely affect the generalizability of the ML-enhanced closure models to unseen flows.

\subsection{\label{sec:Close}Closed loop framework}
\begin{figure}
    \centering
    \includegraphics[width=0.9\textwidth]{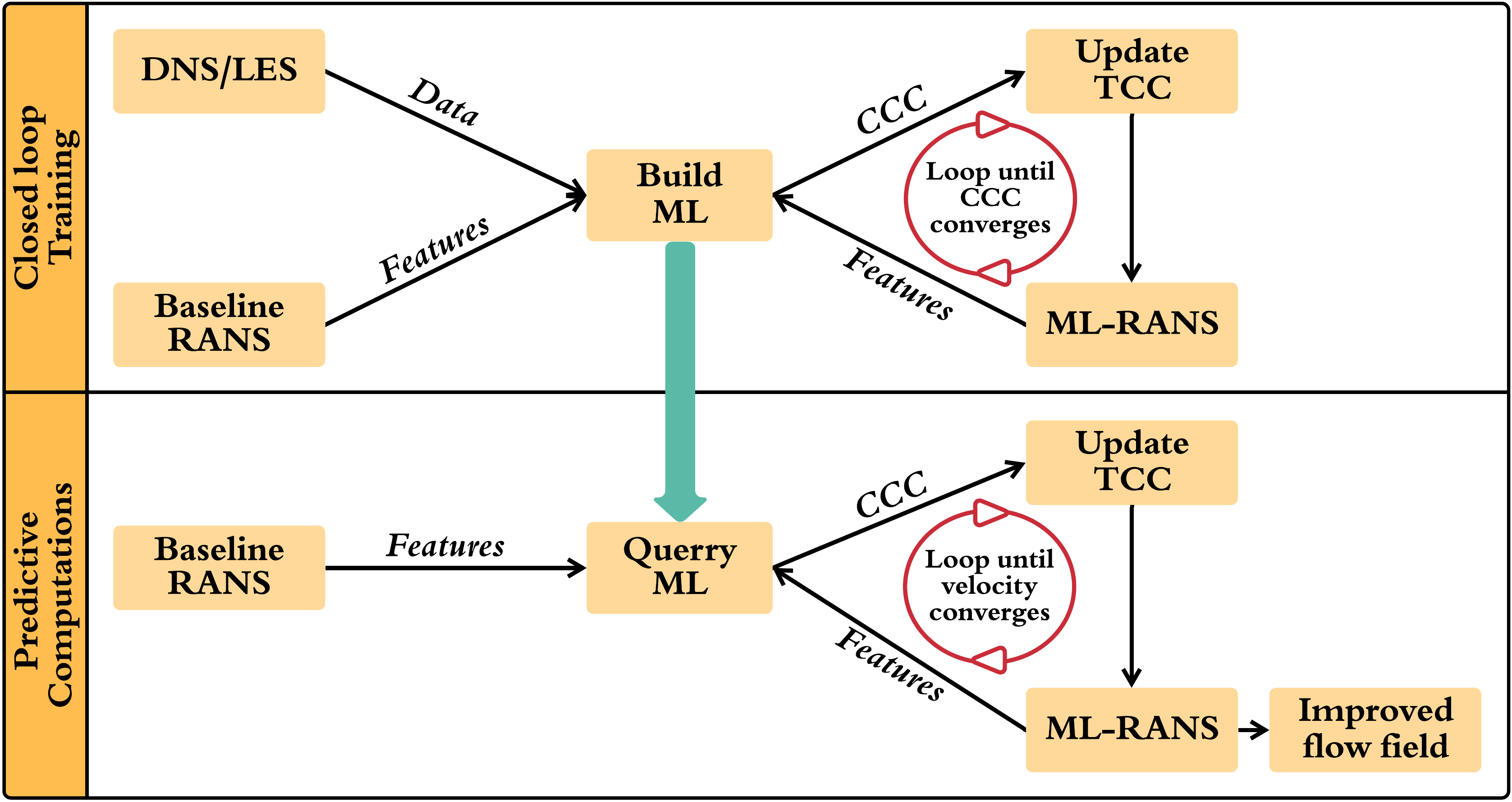}
    \caption{ Closed loop framework}
    \label{Close}
\end{figure}

It is  desirable to incorporate some form of dynamical systems analysis into the ML training process to ensure consistency and compatibility between the various coefficients.  However, due to the implicit nature of the learned functionals, analytical approaches are not straightforward. Instead, we propose embodying some degree of  physical compatibility and training consistency into the modeling process by adopting a \emph{closed loop} training framework. A schematic of one such framework is given in Fig.~\ref{Close}.

\textbf{Training procedure:} The first few steps of the closed loop training procedure are similar to those of the open loop framework. Baseline RANS produces the initial features which are used in conjunction with high fidelity data to build the initial ML functional for values of CCC. In the open loop process, this functional is used directly in predictive computations. However, in the closed loop procedure, several additional steps are involved as shown in the upper schematic in Fig.~\ref{Close}. First, the TCC values are modified (e.g., using Eq.~\eqref{eq:sigma}) for compatibility with new CCC values. If the new and baseline TCC and CCC values are nearly identical (based on some convergence criteria), then the training process is complete. If not, an iterative looping procedure is performed as follows. A RANS computation is performed with the new (Loop-1) TCC and CCC leading to updated values for the input  features for ML training. These updated features and high fidelity data are used to retrain ML functional. In retraining process, the learned parameters (e.g., weights and biases of neurons in neural network technique) of Loop 1 are used to initialize the ML algorithm. This process of reusing and transferring of the prior knowledge is similar to \emph{transfer learning} \cite{weiss2016survey}. Using the retrained CCC values, the TCC values are modified once again using Eq.~\eqref{eq:sigma} for compatibility. If the retrained (Loop-2) and previous (Loop-1) TCC-CCC values are not nearly the same, the looping sequence continues until convergence is achieved. The converged ML functional for CCC and corresponding TCC values are then deemed suitable for use in predictive ML-RANS computations.
Thus, in this closed loop training, the iterative looping process ensures consistency between the system of equations that produces the features and the ML functional that produces the coefficients (CCC and TCC) used in the equations. This consistency amongst the various closure coefficients can lead to improved generalizability in unseen flows.  

\textbf{Predictive computations:} A schematic of the predictive closed loop computation of unseen flows is shown in Fig.~\ref{Close}. First, baseline RANS is performed to provide the initial features - mean flow strain rate, rotation rate, turbulent kinetic energy and dissipation fields. Then the trained ML functional is invoked to iteratively update CCC and TCC  values. Iterations are performed until the flow variables converge to desired levels. With enhanced compatibilty and consistency of the closure coefficients in closed loop framework, it is expected that converged flow field will represent an improvement over the baseline RANS and open loop computations.

The potential shortcomings of open loop training and the proof-of-concept of the closed loop framework are examined next in a simple turbulent flow.

\section{\label{sec:Proof}Proof-of-concept studies}
As highlighted in Sec.~\ref{S:1}, the goal of this study is to develop fundamental guidelines for advancing ML-RANS approach, rather than to propose a specific model. 
In this section, we formulate simple test studies to investigate training concepts discussed in the previous sections.
The objectives of the proof-of-concept studies are to examine:
\emph{(i)} the inconsistencies that can rise from current open loop training framework, 
and \emph{(ii)} the improvements enabled by imposing compatibility constraints and the closed loop training procedure. 
It bears repeating that the goal of this study is to establish foundational guidelines for developing ML-RANS rather than to develop a specific ML-closure. To investigate the concepts presented in the previous sections, we seek answers to the following questions:
\begin{enumerate}[noitemsep]
  \item How can ML-RANS models help in flows in which baseline RANS performs reasonably well?
  \item How does ML-RANS perform in flows in which baseline Reynolds stress constitutive relation is incorrect?
  \item How does ML-RANS perform in flows in which baseline modeled transport equations are inadequate?
\end{enumerate}
Three studies are formulated to address each of the above questions.

The choice of flow for the above demonstration must be made judiciously. 
In many complex flows involving separation and coherent structures, the very paradigm of a Reynolds stress constitutive relation is questionable due to dominant spatial and temporal non-local effects. Thus, the very paradigm of two-equation RANS models may be formally invalid. Therefore, to examine ML training frameworks (open vs. closed loop) a prudent choice would be a simple flow in which the two-equation RANS approach is reasonably valid. Study of such a flow is of value as any ML-RANS that does not perform well in simpler flows will be unsuitable for complex flows.

The turbulent channel flow has served as an important benchmark case for RANS and higher-order closure model development \cite{wilcox1998turbulence}. Many simple two-equation RANS models have been calibrated to yield a good agreement for the mean flow profile and the turbulent shear stress. However, the anisotropy of the turbulent normal stresses are not very well captured if Boussinesq constitutive relation is used. Anisotropic eddy viscosity models can improve the prediction of turbulence normal stresses. 
Nevertheless, the standard $k$-$\omega$ model discussed in Sec.~\ref{sec:RANS} will serve as the baseline model. 
The goal of the first study will be to examine if the ML training process can result in an anisotropic constitutive relation which can yield improved prediction of the normal anisotropy components. 
Rather than use other more complex flows for the second and third studies, we still use the channel flow. For the second study, to emulate the effect of an inadequate Reynolds stress constitutive relation, 
the standard CCC values are modified to unphysical values and the standard TCC values are retained intact. Thus the baseline RANS model used in the training process is an intentionally degraded $k$-$\omega$ model.
The aim of the second study is to examine if the ML training process will recover the correct CCC values.
For the third study, standard TCC will be modified but CCC values retained intact from the standard model. Here, we will determine if the training will lead to reasonable predictions.
If the ML training technique is adequate, then in second and third studies, the ML-RANS should overcome the incorrectly initialized coefficients to recover the correct values and yield accurate results. 

\subsection{\label{sec:Obj} Objective function definition}
The success of the ML-RANS depends on the choice of the objective function used for optimization of the 
coefficients
during the learning process. In this study we seek a ML-functional for the Reynolds stress constitutive relationship. Thus there are two choices of labels for defining the objective function --  normalized anisotropy tensor $(b_{ij})$ or Reynolds stress tensor $(\langle u_iu_j\rangle$).  

If the objective function in ML algorithm is based on normalized anisotropy tensor $(b_{ij})$ \cite{ling2016reynolds}, ML-RANS turbulence model can be expected to reproduce normal anisotropy components in the channel flow
adequately. 
However, since RANS model does not necessarily predict the accurate turbulent kinetic energy $(k)$, the final Reynolds stress tensor $(\langle u_iu_j\rangle)$ will not be accurate.
Incorrect Reynolds shear stress leads to wrong mean velocity profile and friction velocity. Indeed, the all important `log-layer' features may also not be accurately captured. Thus, casting the objective function exclusively in terms of anisotropy tensor can lead to errors in computing important flow quantities.

Constructing the objective function in terms of Reynolds stress tensor $(\langle u_iu_j\rangle)$ \cite{geneva2019quantifying} is the other option. This will certainly lead to an adequate computation of the mean velocity field. However, this can lead to another important inconsistency.  As mentioned earlier, the DNS and RANS kinetic energy can be quite different. Thus the ML-functional for turbulent kinetic energy $(k)$ and the value of $k$ obtained from the modeled transport equation will not be the same leading to a disparity in the computed results.
It is evident that some degree of disparity in the computed results is inevitable when ML-functionals and modeled transport equations are used in combination to simulate turbulent flows.

The objective function must be constructed based on the desired features of the computed flow. Here, we designate the following features as the required elements of ML-RANS computation:
\begin{enumerate}[noitemsep]
  \item Accurate log-law velocity profile,
  \item Accurate Reynolds shear stress $(\langle u_1u_2 \rangle)$,
  \item Maintain the following equality:
  
  $\langle u_1u_1\rangle+\langle u_2u_2\rangle+\langle u_3u_3\rangle$ (ML-RANS) = $2k$ (RANS transport equations), 
  \item Preserve realizability: $\langle u_\alpha u_\alpha\rangle \geq 0$.
\end{enumerate}
Towards this end, the objective or mean square error (MSE) loss function for ML training is defined as,
\begin{equation}
MSE=\frac{1}{4N}\sum_{m=1}^{N}\Big[\sum_{\alpha=1}^{3} (b_{\alpha \alpha}-\widehat{b}_{\alpha \alpha})^2+\frac{1}{u^4_{\tau}}(\langle u_1u_2\rangle-\widehat{\langle u_1u_2\rangle})^2\Big],
\label{eq:LossFunction}
\end{equation}
where predicted outputs of the ML algorithm are denoted by $\widehat{b}_{\alpha \alpha}$, $\widehat{\langle u_1u_2\rangle}$ and the true DNS values are shown by ${b}_{\alpha \alpha}$ and $\langle u_1u_2\rangle$. Here, $N$ represents the number of data points. In this definitions, shear stress component is normalized by true DNS value of the friction velocity ($u_{\tau}$) to ensure a consistent velocity. 
Therefore, for the choice of flow considered in this study, i.e., channel flow, the ML objective function is defined based on Reynolds shear stress and normal anisotropy components.


Models based on fundamental physical principles can be expected to yield reasonable results for quantities not invoked in the coefficient calibration process. One of the limitations of data-based methods is that the accuracy of quantities not involved in the loss (objective) function is unclear. In the
computations we will examine the ability of the ML-RANS models to predict important flow quantities not used in the ML loss function: production-dissipation ratio $(P/\epsilon)$ and mean flow to turbulence frequency $(Sk/\epsilon)$.

In summary, while turbulent channel flow is a simple benchmark problem, it is ideally suited for examining important concepts on how ML can be used to enhance RANS.

\subsection{\label{sec:Net} Neural networks}

\begin{figure}
    \centering
    \includegraphics[width=0.7\textwidth]{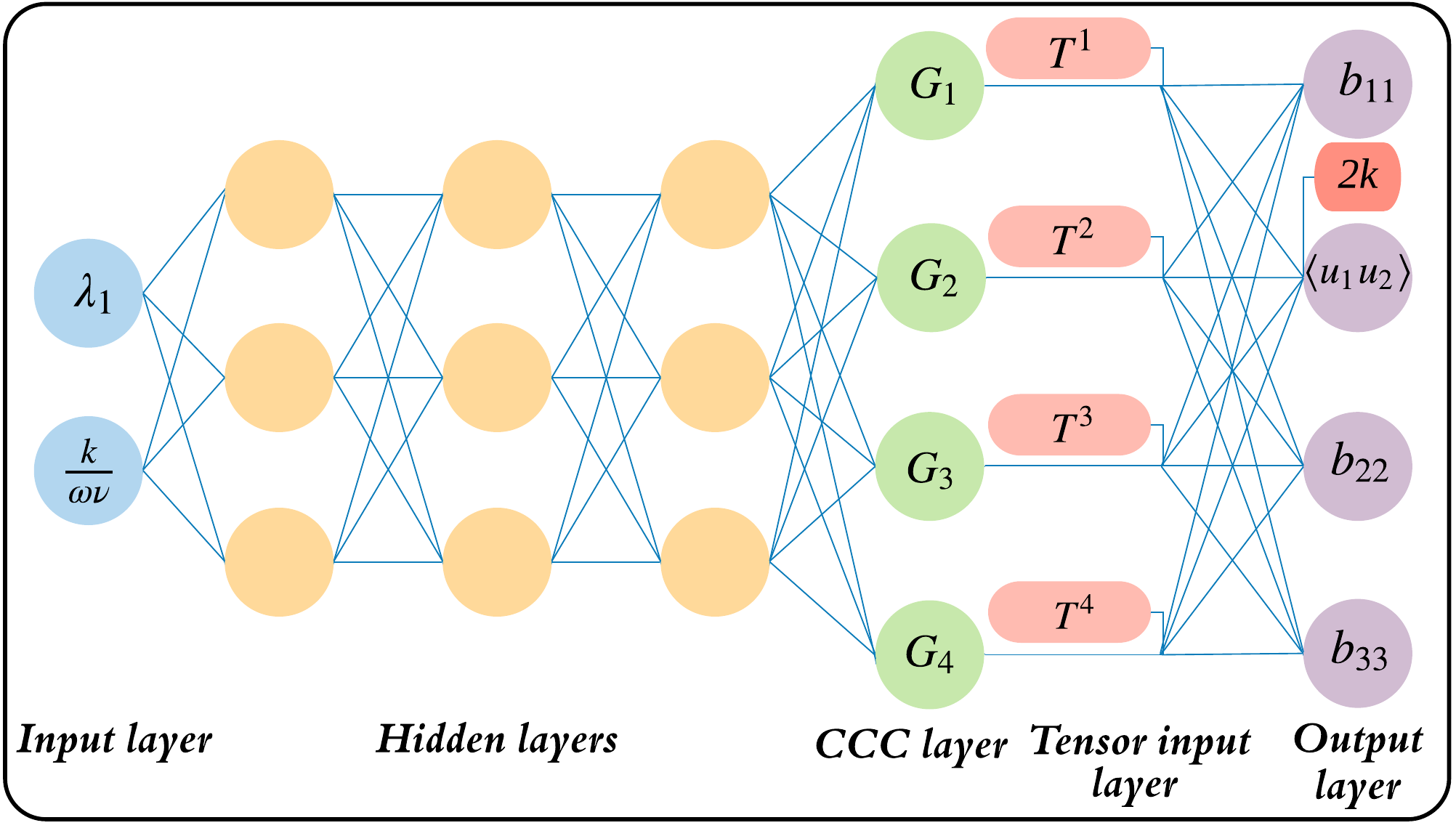}
    \caption{ Schematic of the fully connected feed-forward neural network}
    \label{NN}
\end{figure}

Recently, various ML algorithms including neural networks and random forests have been used for modeling fluid dynamics in general and turbulence in particular. Neural networks have been shown to have superior performance in modeling non-linear and complicated relationships with high-dimensional data \cite{ling2016reynolds}. It has been shown that incorporating Galilean invariant features further enhances the generalizability of the neural networks \cite{ling2016machine}. 
In this study Tensorflow \cite{abadi2016tensorflow}, which is widely used and comparatively well documented library, is employed for ML computations. A fully connected feed-forward neural network is trained using backpropagation with gradient descent method.
The schematic of the selected neural network architecture is shown in Fig.~\ref{NN}. To control the overfitting of the neural network during the training, a L2-norm regularization term is imposed on the loss function (also known as Ridge-regression \cite{murphy2012machine}) to constrain the magnitudes of the learning parameters (weights and biases of neurons). A grid search approach is adopted for hyperparameter optimization. The details of the optimized network architecture used in this work are given in Table \ref{tab:1}. The neural network library, Tensorflow is linked to open source CFD code OpenFOAM \cite{weller1998tensorial} using the provided C API.

\begin{table}
  \caption{Neural network hyperparameters}
  \label{tab:1}
  \begin{tabularx}{0.8\textwidth}{XX}
    \hline\hline
    Name & Value\\
    \hline
    Number of hidden layers & 3 \\
    Number of nodes per layer & 3 \\
    Activation function & Elu \\
    Optimization algorithm & Adam \cite{kingma2014adam}\\
    L2-norm penalization coefficient & $7\times10^{-3}$\\
    Learning rate & $1\times10^{-4}$ \\
    Initialization function & Xavier normal \\    
    \hline\hline
  \end{tabularx}
\end{table}

\section{\label{sec:Find}Results}
The computations in this section are aimed at highlighting the training stage of closed loop method and contrasting the difference between open and closed loop frameworks in predictive computations. The demonstration is performed in a turbulent channel flow which is one of the simplest non-trivial cases of closure modeling interest. With this flow choice the inherent limitations of two-equation RANS closure paradigm that affect more complicated cases do not obfuscate the training inadequacies. The standard two-equation $k-\omega$ model is used without near-wall low Reynolds number corrections. The low Reynolds number corrections are precluded, as one of the goals of the study is to determine if the ML training can enable the Reynolds stress constitutive model to capture these effects. The transport and constitutive equations describing the standard model are given in Sec.~\ref{sec:RANS} and further details are available in \cite{wilcox1998turbulence}.

Given that the channel flow is statistically two-dimensional, the constitutive equation we seek to train in this study requires only four basis tensors \cite{pope2001turbulent},
\begin{equation}  
b_{ij} = G_1(s_{ij})+G_2(s_{ik}r_{kj}-r_{ik}s_{kj})+G_3(s_{ik}s_{kj}-\frac{1}{3}\delta_{ij}s_{mn}s_{nm})+G_4(r_{ik}r_{kj}-\frac{1}{3}\delta_{ij}r_{mn}r_{nm}).
\label{eq:Bous}
\end{equation}  
Consideration is restricted to two scalar input features, $\lambda_1$ and $\frac{k}{\omega \nu}$  to determine the CCC from ML functional, i.e., $G_{n}=g^{n}(\lambda_1, \frac{k}{\omega \nu})$.

Datasets used in training, validation and predictive calculations are shown in Table \ref{tab:2}. For training purposes, DNS data \cite{lee2015direct} within the wall-normal distance range $0<y/h<0.8$ is employed, where $h$ is the channel  half-width. The points in the regions near the channel center ($y/h>0.8$) are excluded to prevent the model coefficients from becoming unphysical as stresses tend to zero \cite{sotgiu2019towards}. 

We examine the use of two compatibility constraints. The relationship given in Eq.~\eqref{eq:sigma} is used in baseline models and all of the closed loop training procedures. Not imposing this constraint leads to poor ML-RANS behavior in the log-layer of the channel. The effect of using constraint implied by Eq.~\eqref{eq:ProductionDissipation} is also investigated. 
\begin{table}
  \caption{Selected datasets}
  \label{tab:2}
  \begin{tabularx}{\textwidth}{XXX}
    \hline\hline
    $Re_\tau $ & $Re_{b}$ & Purpose \\
    \hline
    550 & 10,000 & validation of the ML algorithm  \\
    1000 & 20,000 & training of the ML algorithm\\
    5200 & 125,000 & predictive computation \\
    \hline\hline
  \end{tabularx}
\end{table}
\subsection{\label{sec:training} Investigation of the closed loop training approach}
As mentioned in Sec.~\ref{sec:Proof}, training is performed for three different scenarios to investigate various aspects of ML-RANS predictive capabilities. Details of the baseline models employed in the three studies are shown in Table \ref{tab:3}. Here we use the term \emph{baseline} to describe the RANS model which initiates the ML procedure. The computations of each study are directed toward answering three questions:
\begin{enumerate}[noitemsep]
  \item To what extent are the standard values of \emph{G's} recovered by ML when they are intentionally altered in baseline model?
  \item Do the trained ML values of \emph{G’s} lead to marked improvement in anisotropy predictions?
  \item How well does ML-RANS perform toward capturing quantities of interest not included in definition of the ML loss function?
\end{enumerate}

\begin{table}
  \caption{Baseline model coefficients used in different studies}
  \label{tab:3}
  \begin{tabularx}{\textwidth}{XXXXXXXXXX}
    \hline\hline
      & $\alpha$ & $\beta$ & $\beta^*$ & $\sigma$ & $\sigma^*$ & $G_1$ & $G_2$ & $G_3$ & $G_4$\\
    \hline
    Case-1 & 0.52  &  0.072  & 0.09 & 0.5 & 0.5 & -0.09 & 0 & 0 & 0 \\
    Case-2 & 0.52 & 0.072 & 0.09 & 0.23 & 0.23 & -0.045 & 0 & 0 & 0 \\
    Case-3 & 0.52 & 0.054 & 0.09 & 0.143 & 0.143 & -0.09 & 0 & 0 & 0 \\
    \hline\hline
  \end{tabularx}
\end{table}
\subsubsection{\label{sec:case1} Case-1: Standard $k-\omega$ baseline model}
\begin{figure}
    \centering
    \includegraphics[page=1]{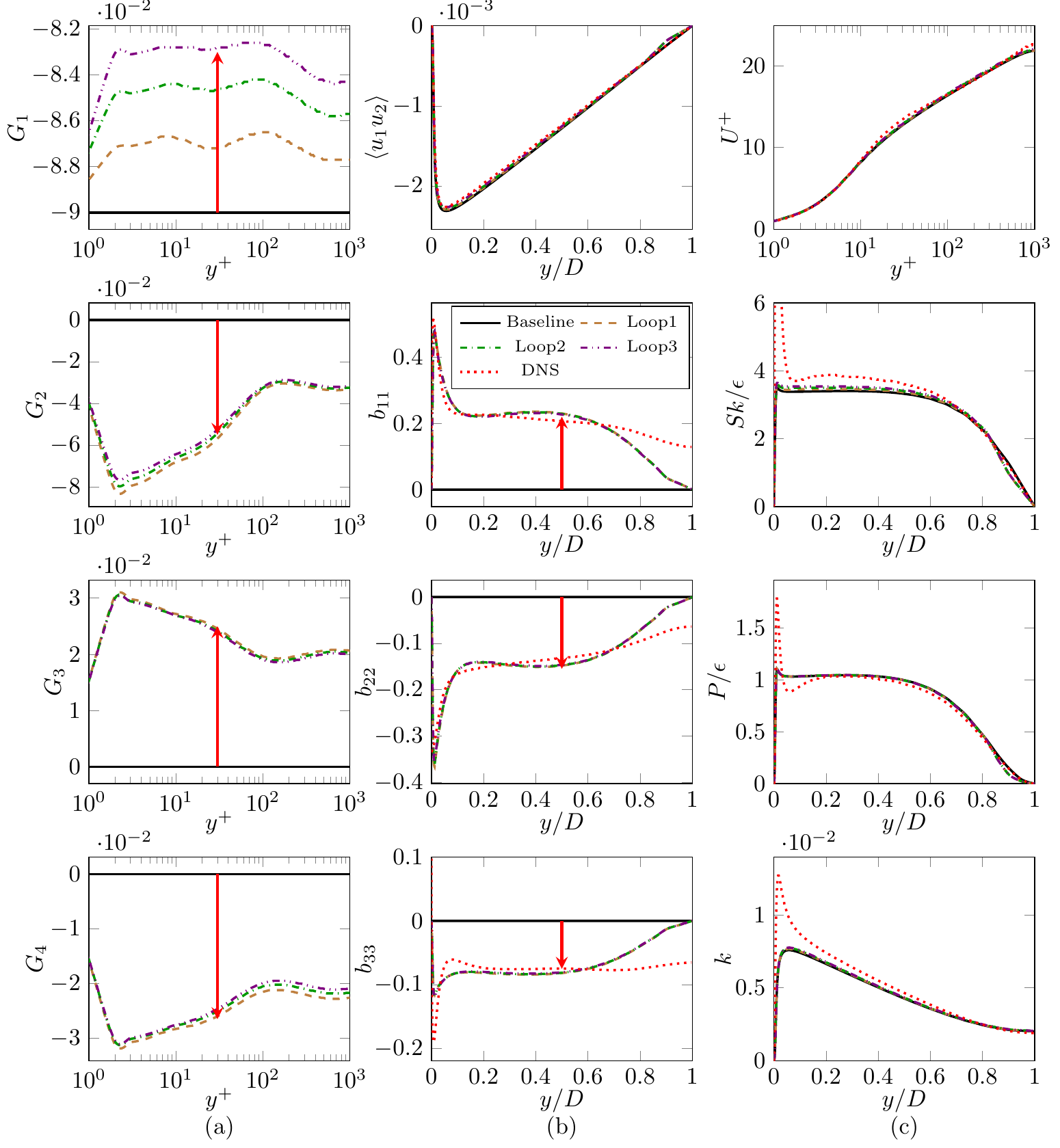}
    \caption{Closed loop training for Case-1, (a) CCC, (b) turbulent shear stress and normal anisotropy components, (c) other QoI. The red arrow indicates the direction of increasing training loops.}
    \label{S1-Main}
\end{figure}

In this study the standard $k-\omega$ closure serves as the baseline model. The model is then trained with closed loop procedure. Computed results obtained by the baseline and ML-RANS simulations are compared with DNS in Fig.~\ref{S1-Main}. It is seen that by training the ML algorithm on multiple loops, the magnitude of the $G_1$ coefficient gradually reduces and finally converges to a value around $0.083$. Some variation of $G_1$ on wall-normal distance is seen. Other coefficients converge to non-zero functions. Marked improvements in computing normal components of anisotrpy tensor are observed in ML-RANS simulations and turbulent shear stress is accurately reproduced. The results of the baseline and ML-RANS simulations for other quantities of interest (QoI), i.e., turbulent kinetic energy ($k$), mean flow to turbulence frequency $(Sk/\epsilon)$ and production-dissipation ratio $(P/\epsilon)$ are nearly identical and no significant improvements are observed near the wall with ML-RANS. This is to be expected as the baseline (standard) $k-\omega$  model is tuned and calibrated to yield good agreement of these QoI.

\subsubsection{\label{sec:case2}Case-2: Modified CCC baseline model}
In many applications of interest, the baseline Reynolds stress constitutive equation can be quite inadequate due to incorrect values of CCC.
For such cases, it is expected that ML training with high fidelity data can lead to a ML functional for CCC that is significantly more accurate. Clearly, it is important to establish the capability of ML training procedure to recover from a poor baseline model. Rather than seek a flow in which the standard model is not correct, we simulate the scenario by intentionally modifying the standard Boussinesq model coefficient ($G_1$). 

\begin{figure}
    \centering
    \includegraphics[page=1]{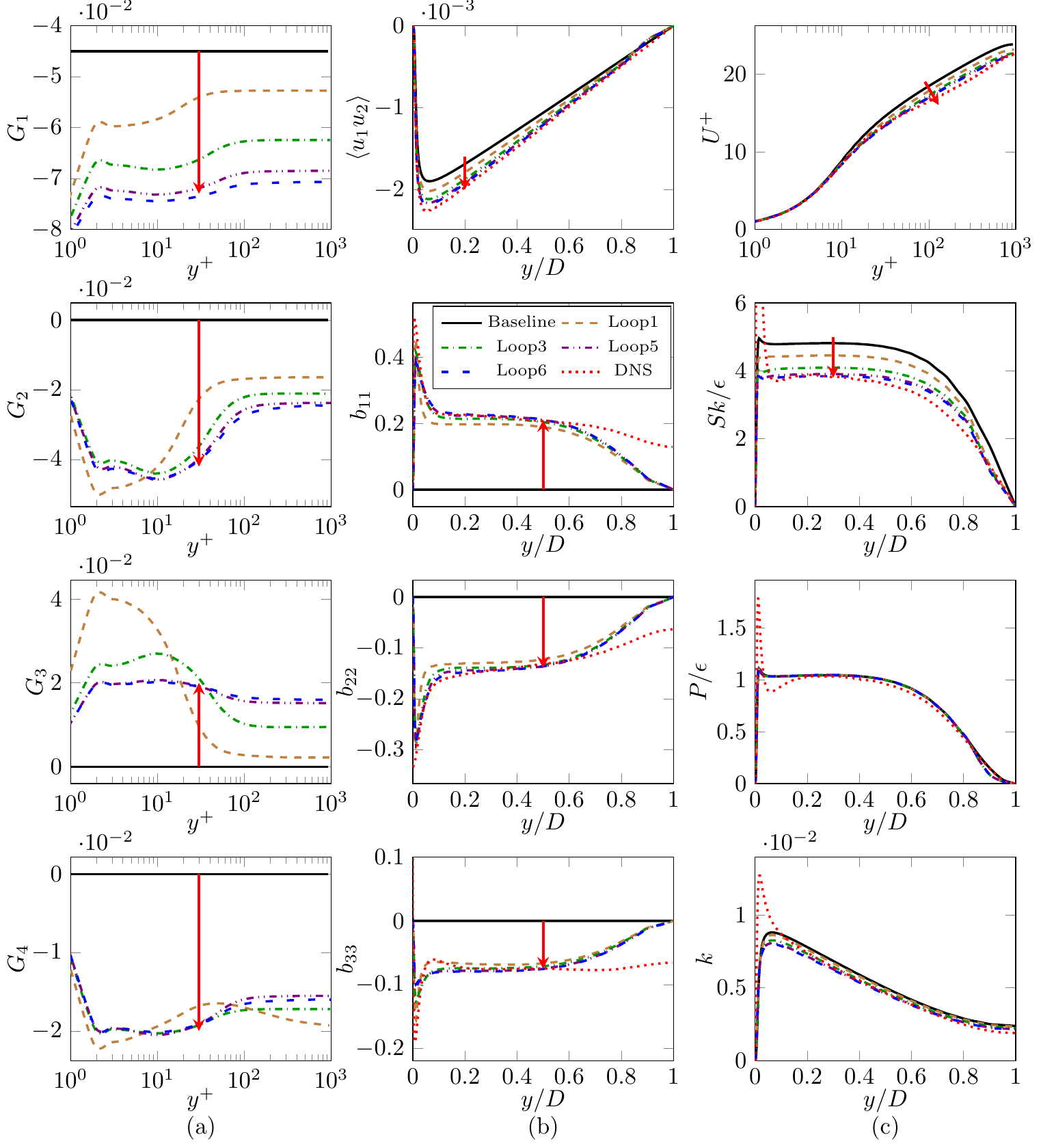}
    \caption{Closed loop training for Case-2, (a) CCC, (b) turbulent shear stress and normal anisotropy components, (c) other QoI.}
    \label{S2-Main}
\end{figure}

Computation results of the baseline and closed loop ML-RANS for this case are presented in Fig.~\ref{S2-Main}. It can be seen that the baseline RANS simulation with modified CCC is inaccurate  for most of the QoI in turbulent channel flow. By performing multiple loops of training, initially incorrect $G_1$ coefficient (= -0.045) used in baseline model recovers to a more reasonable value $G_1 \sim  -0.074$. Other CCC converge to non-zero values similar to those in Case-1. 
The  ML-RANS leads to improved computations of turbulent shear stress and normal anisotropy components. Other QoI such as mean velocity, $Sk/\epsilon$ and $P/\epsilon$ that have not been used in definition of the ML loss function are also  significantly better than the baseline case. It should be noted that the TCC constraint for $\sigma$ is imposed here. The results without this constraint are significantly worse. Thus  the approach of (i) enforcing TCC constraint and (ii) closed loop training leads to recovery from an inaccurate baseline model.   
\subsubsection{\label{sec:case3}Case-3: Modified TCC baseline model}
In many applications, the coefficients in the modeled transport equations (TCC) can be inaccurate. To simulate this effect, in this case a key TCC value ($\beta$) is modified from the standard value as shown in Table \ref{tab:3}. This case is of interest as the incorrect coefficient is not modified with ML training process. Nonetheless, such a scenario can occur in a practical application.

\begin{figure}
    \centering
    \includegraphics[page=1]{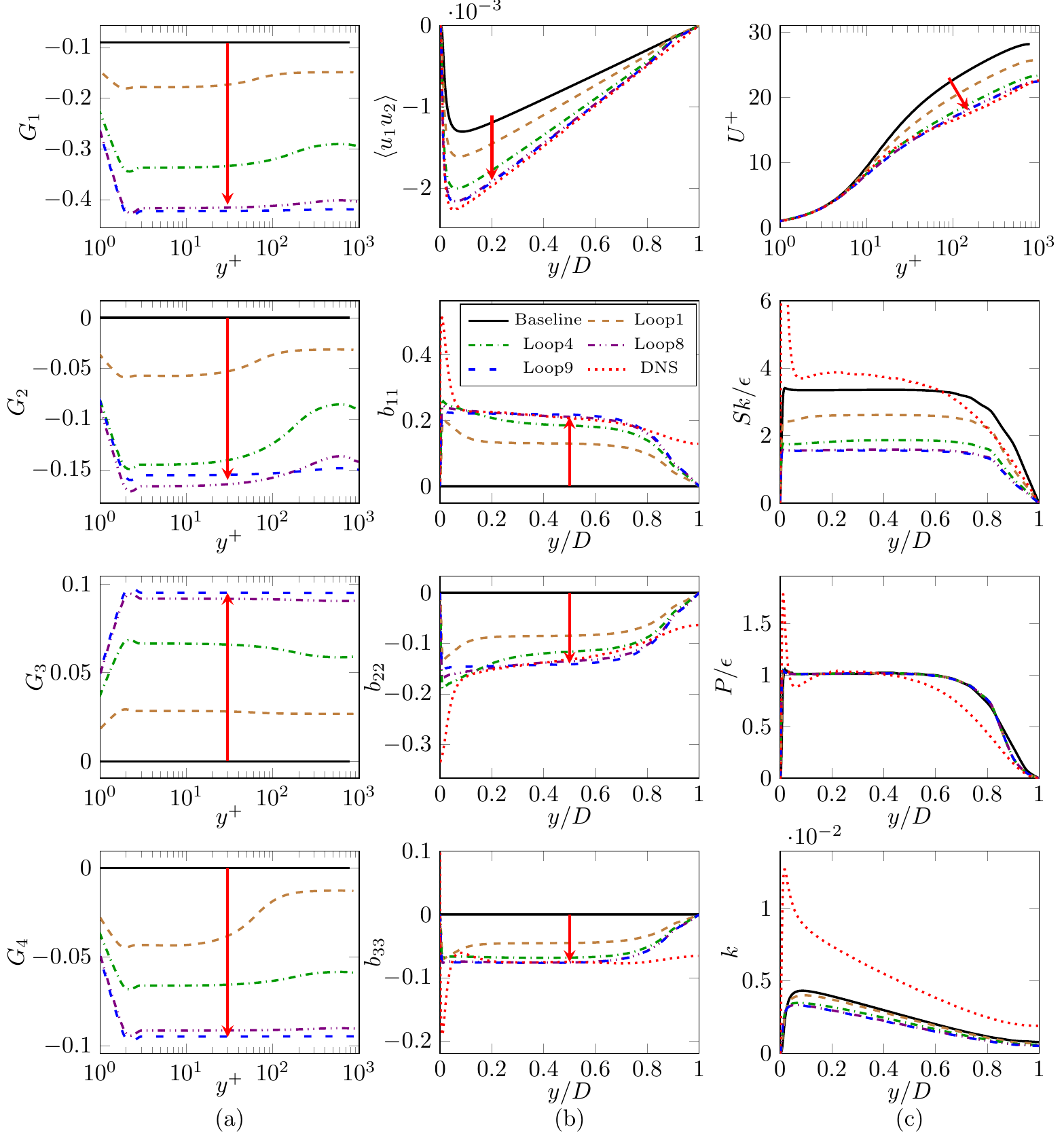}
    \caption{Closed loop training for Case-3, (a) CCC, (b) turbulent shear stress and normal anisotropy components, (c) other QoI.}
    \label{S3-Main}
\end{figure}

The computed results of various quantities by the baseline and the closed loop ML-RANS simulations are compared against DNS data in Fig.~\ref{S3-Main}. To compensate for the altered TCC, ML process changes CCC away from their correct values. For instance, the  value of $G_1$ drifts away from the `correct value'  of about $-0.09$  to about $-0.42$. 
Despite the incorrect CCC, the anisotrpoies and turbulent shear stress are captured reasonably well. Furthermore, the mean velocity and $P/\epsilon$ profiles are also adequately computed. It must be noted that these QoI are directly related to the ML loss function. This exhibits the strength of the closed loop training.  Despite physically incompatible closure model coefficients, the training process provides reasonable prediction of QoI included in the optimization process. The incompatibility of the closure coefficients leads to poor predictions of other quantities such as 
 $Sk/\epsilon$ and \emph{k}. Thus, one of the important challenges in ML-RANS is to ensure reasonable behavior of QoI not related to the ML loss function. In the next subsection, we demonstrate that the behavior can be improved by imposing further TCC constraints. 
 
\subsection{Open loop vs. closed loop training}
The purpose of this set of computations is to compare and contrast open loop and closed loop training procedures. Again, for the sake of simplicity and clear illustration, we use a channel flow at high Reynolds number as the test flow.  Any difference between the two training procedures exhibited in this canonical flow will also manifest in flows of practical interest. 
ML models  trained  at $Re_{\tau}=1000$  are used to perform predictive simulations of channel flow at $Re_{\tau}=5200$ (Table \ref{tab:1}). Three baseline models introduced in Subsec.~\ref{sec:training} are  trained with open and closed loop procedures. In these computations, one or both of the physics-based realizability constraints are  used along with closed loop training.  

\subsubsection{\label{sec:pcase1} Case-1: Standard $k-\omega$ baseline model}
\begin{figure}
    \centering
    \includegraphics[page=1]{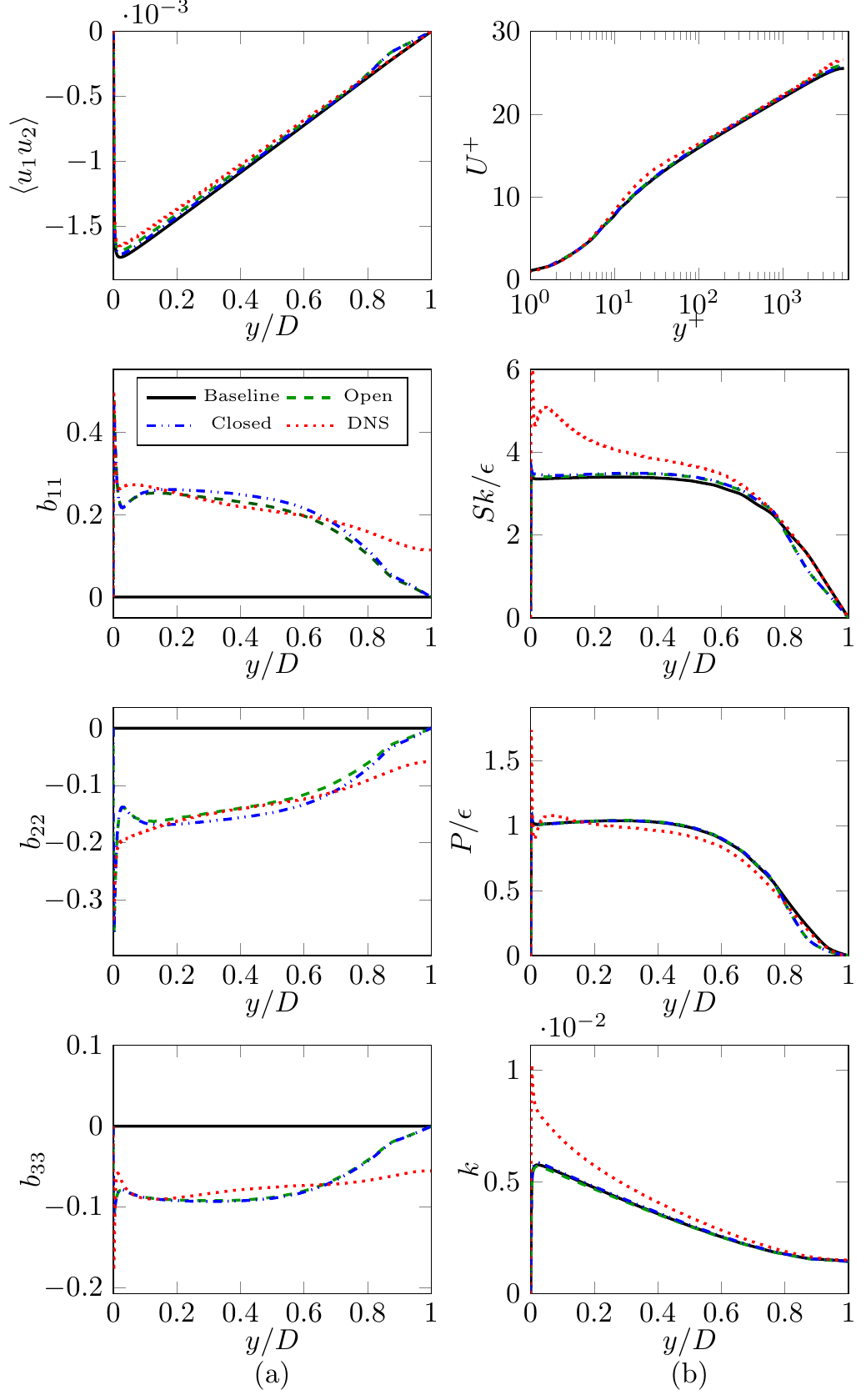}
    \caption{Predictive computations using open and closed loop frameworks for Case-1, (a) turbulent shear stress and normal anisotropy components, (b) other QoI.}
    \label{S1-Predict}
\end{figure}

The computed results of the ML-RANS with open and closed loop models are plotted against baseline RANS and DNS data in Fig.~\ref{S1-Predict}. As expected, the baseline RANS simulation with standard $k-\omega$ model reasonably computes most of the QoI in channel flow even at this higher Reynolds number. However, the Reynolds stress anisotropies are not well captured as the baseline model employs the isotropic Boussinesq constitutive relation. Both open and closed loop training result in nonlinear anisotropic constitutive relations.
This  leads to significant improvement in computation of normal anisotropy components with both types of ML training.
For this case which the baseline model accurately computes most of the QoI, nearly identical results are computed by open and closed loop methods.

\subsubsection{\label{sec:pcase2}Case-2: Modified CCC baseline model}
\begin{figure}
    \centering
    \includegraphics[page=1]{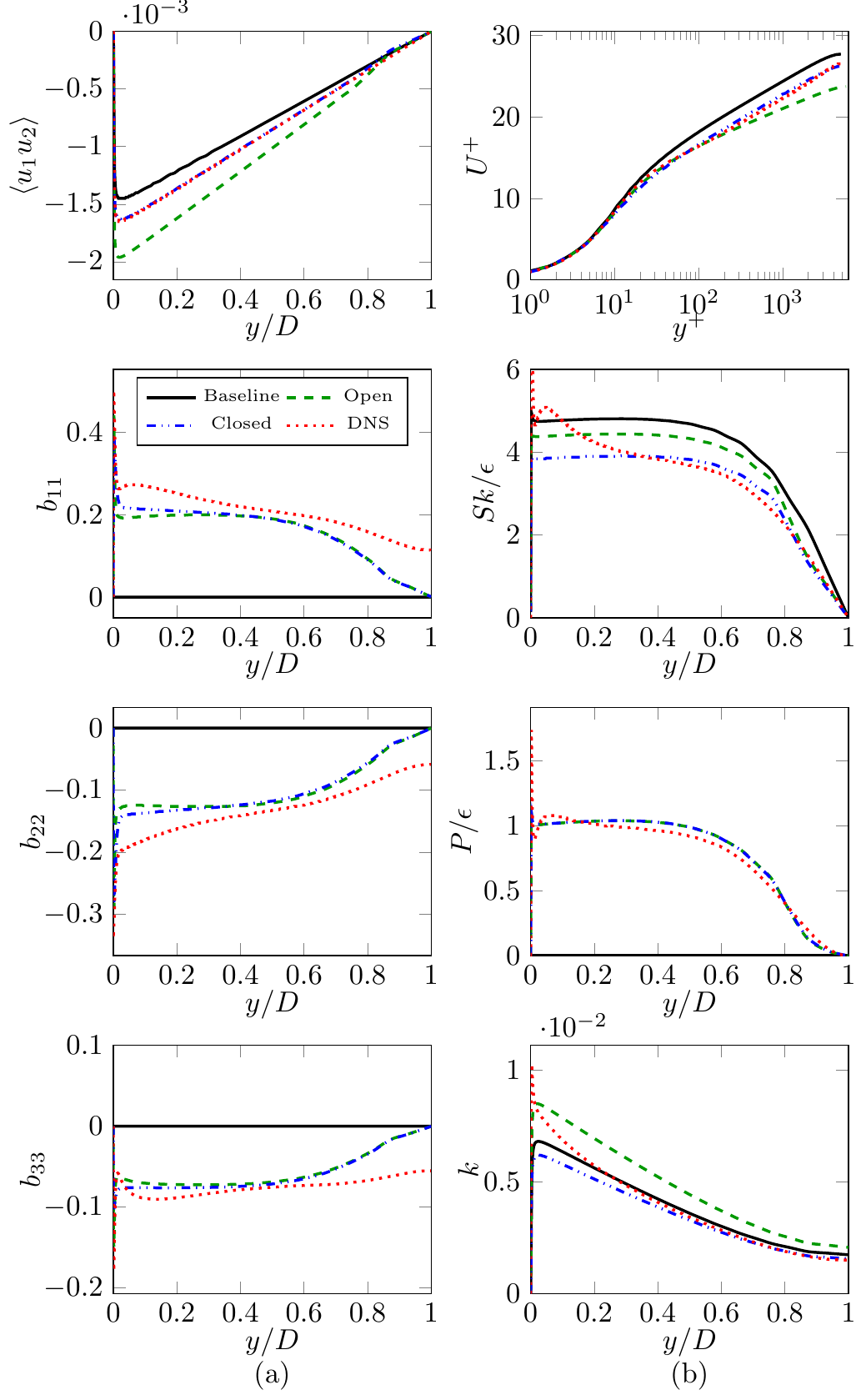}
    \caption{Predictive computations using open and closed loop frameworks for Case-2, (a) turbulent shear stress and normal anisotropy components, (b) other QoI.}
    \label{S2-Predict}
\end{figure}

The results of the ML-RANS and (modified) baseline models are compared with DNS data for channel flow at $Re_{\tau}=5200$ in Fig.~\ref{S2-Predict}. 
As expected, the baseline model results are poor for most QoI. 
The magnitude of turbulent shear stress is significantly lower for the baseline RANS simulation due to the lower $G_1$ value. In turn, this leads to steeper growth of the  mean velocity in log-layer.  Due to the steeper mean velocity gradient, $Sk/\epsilon$ is also high in most parts of the channel. Next, we examine the open loop model results. The one-step training in this case over-corrects the $G_1$ value as illustrated by the larger magnitudes of Reynolds shear stress $(\langle u_1u_2\rangle)$. As a consequence, the mean velocity gradient in the log-layer is less steep than the DNS profile.  The various anisotropies are better captured than in the baseline case. Finally, we investigate the closed loop model. Due to the multi-step training procedure, the  value of $G_1$ is reasonably accurate. 
This is reflected in the precise computation of Reynolds shear stress, mean velocity  and $Sk/\epsilon$ profiles. The normal stress anisotropies are  also well captured. 
The production-to-dissipation ratio $(P/\epsilon)$ is reasonably captured  with all models due to the imposition of the $\sigma$ constraint given in Eq.~\eqref{eq:sigma}. Evidently, the closed loop training leads to markedly improved results.
\subsubsection{\label{sec:pcase3}Case-3: Modified TCC baseline model}
\begin{figure}
    \centering
    \includegraphics[page=1]{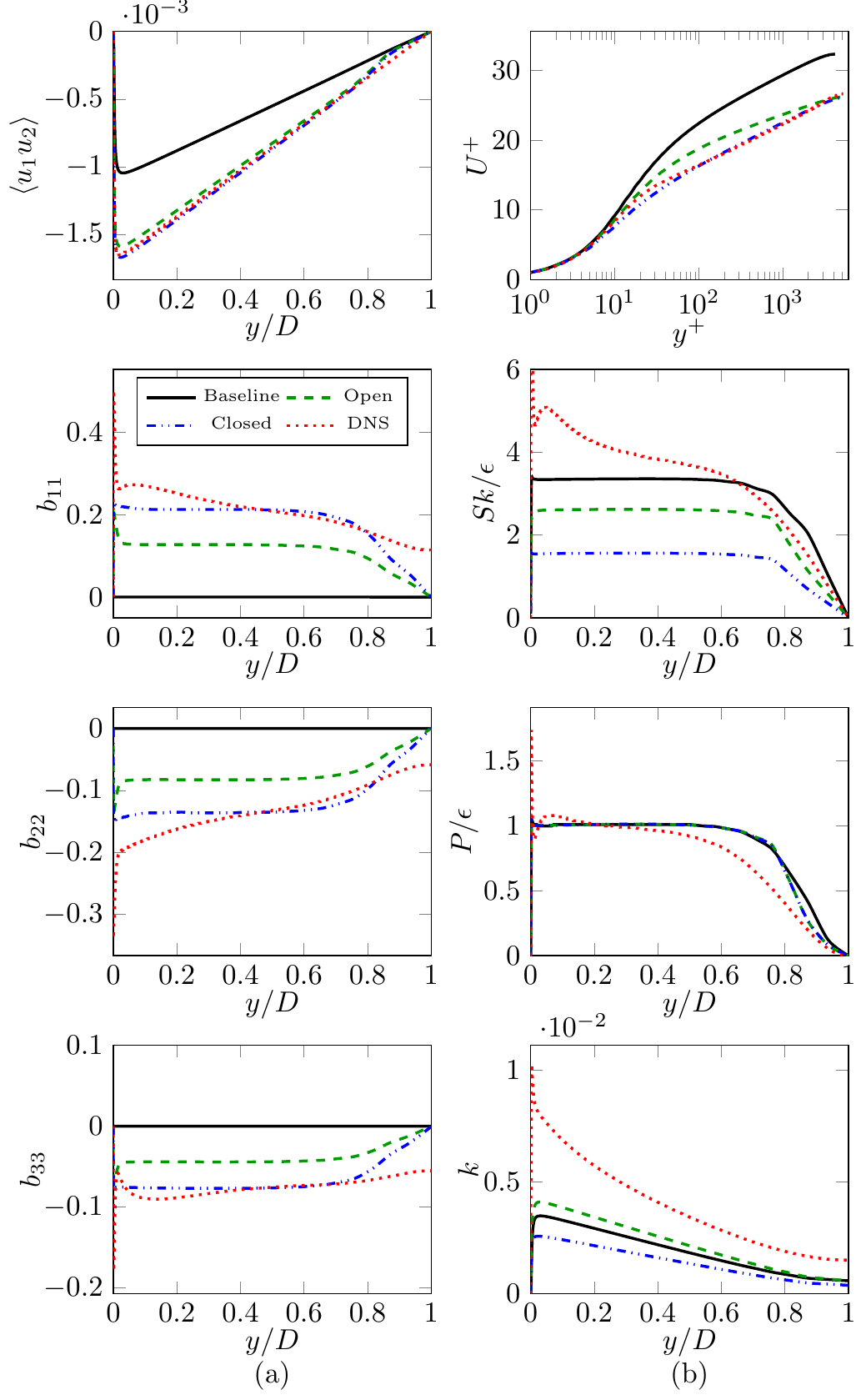}
    \caption{Predictive computations using open and closed loop frameworks for Case-3, (a) turbulent shear stress and normal anisotropy components, (b) other QoI}
    \label{S3-Predict}
\end{figure}

Here we present results for the baseline RANS model with modified TCC and standard CCC values. The results obtained with this baseline and ML-RANS models with different training approaches along with DNS are presented in Fig.~\ref{S3-Predict}. It is seen that baseline RANS severely underpredicts the magnitude of turbulent shear stress. As a result, the mean velocity profile exhibits extended buffer region and late onset of log-law behavior.  The kinetic energy level is much smaller than DNS. The open loop ML-RANS shows significant improvement in Reynolds shear stress and the mean velocity profile moves closer to the DNS case. The magnitude of normal anisotropy components are still low and $Sk/\epsilon$ behavior is worse than the baseline case.  The closed loop ML-RANS shows the best agreement with DNS for Reynolds shear stress and normal stress anisotropy. The mean velocity profile exhibits excellent match with DNS. However, for kinetic energy and $Sk/\epsilon$, the agreement is poor. The closed loop training exhibits the best prediction characteristics for quantities included in the ML loss function. However, this improved behavior comes at the cost of  poor behavior of QoI not included in the loss function. 

The above findings lead to the question whether additional physical compatibility constraints can improve the predictive capability of the closed loop models.
\subsubsection{\label{sec:pcase4} Two compatibility constraints}
\begin{figure}
    \centering
    \includegraphics[page=1]{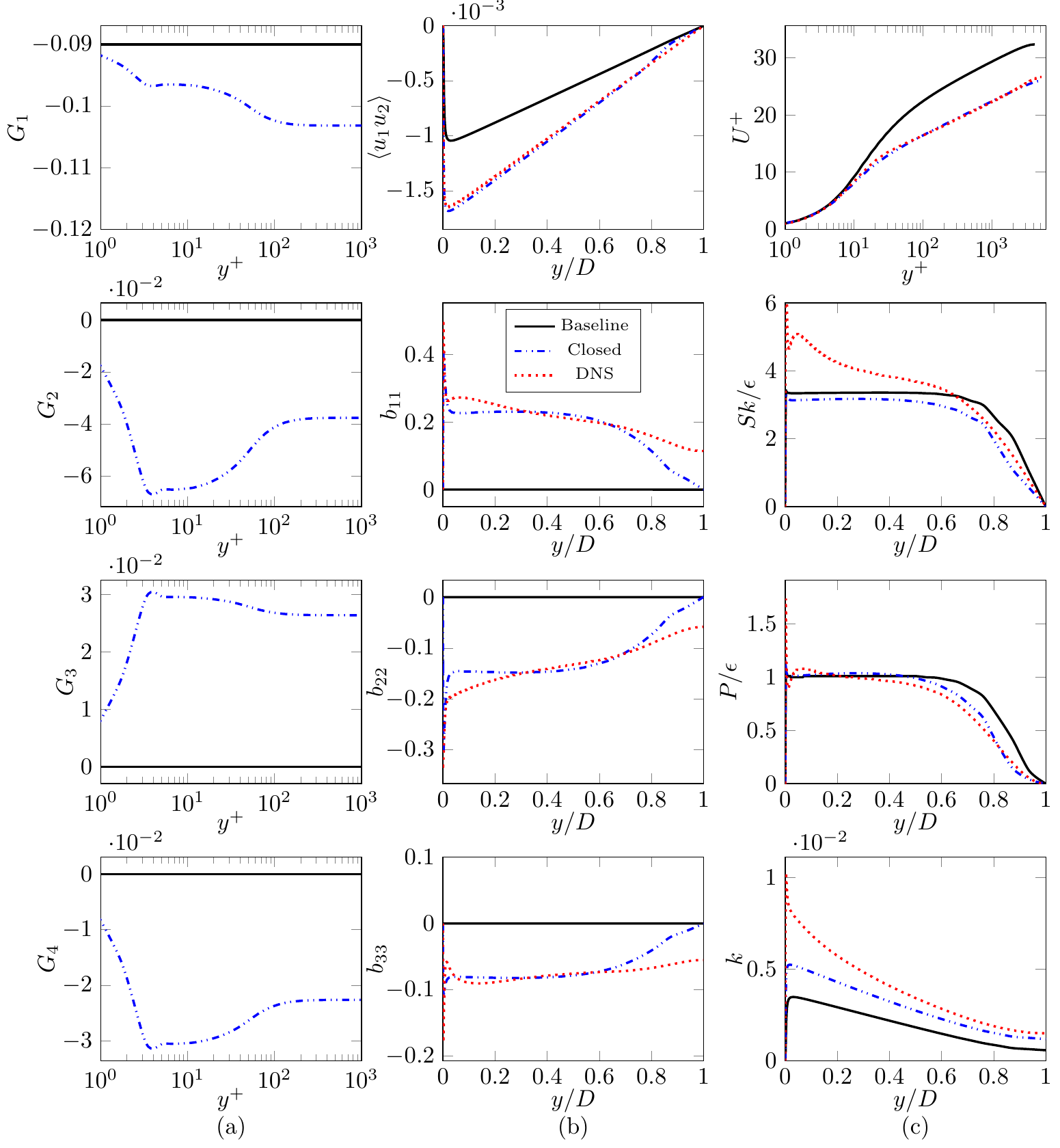}
    \caption{Predictive computations using closed loop frameworks for Case-3 with two-constraining conditions, (a) CCC, (b) turbulent shear stress and normal anisotropy components, (c) other QoI.}
    \label{S4-Predict}
\end{figure}

In all of the closed loop ML-RANS results presented so far, only one CCC-TCC compatibility constraint (Eq.~\eqref{eq:sigma}) was enforced. As mentioned earlier, this constraint is critically important for two reasons: (i) obtaining the correct log-layer slope; and (ii) yielding $P = \epsilon$ in the log-layer. We now perform training  by imposing a second constraint from Eq.~\eqref{eq:ProductionDissipation}.  Case-3 serves as the baseline model for this study.  The results from the baseline RANS and ML-RANS (closed loop) are compared with DNS in Fig.~\ref{S4-Predict}. The converged closed loop trained CCC profiles are exhibited in the first column of the figure. The $G_1$ values produced by two-constraint training are about $10\%$ higher in magnitude than the standard value of $0.09$. Note that in the training studies shown in the previous subsection, the $G_1$ values for Case-1 and Case-2 were about $10-15\%$ lower in magnitude. The other $G$ values in this two-constraint training are quite close to those in Case-1 and Case-2, Figs.~\ref{S1-Main} and \ref{S2-Main}. The Reynolds shear stress and normal anisotropies from the two-constraint ML-RANS are in good agreement with DNS data. The mean velocity profile is  well captured. The production-to-dissipation ratio is also in reasonable agreement with data. This is to be expected as a consequence of enforcement of the first compatibility constraint. The benefit of imposing the second compatibility constraint is evident from the profile of  $Sk/\epsilon$. There is a significant improvement in the predicted profile compared to the Case-3 one-constraint training. 

Overall, the results presented in this section provides evidence of the importance of  (i) closed loop training;  and (ii)  imposition of appropriate constraints.  Although the demonstration has been provided only in the case of the simple channel flow, internal consistency (closed loop) and physical compatibility will be even more important in complex engineering flows.

\section{\label{sec:Conclusion}Summary and Conclusion}
Turbulence models which incorporate machine learning (ML) techniques in the closure scheme have the potential to transform the computation of complex flows of practical interest. In principle, ML-enhancement can improve modeling capabilities at all closure levels ranging from RANS equations to LES. ML-enhanced RANS is of particular interest as it can have an immediate impact on complex engineering applications. To maximize the utility of ML on turbulence closures, it is important to formulate foundational tenets to guide ML-enhanced model development. This study proposes fundamental guidelines for incorporating ML-elements into two-equation RANS closures.
The traditional (physics-based) RANS model equations constitute a dynamical system, wherein the closure coefficients are carefully calibrated to yield reasonable results in a set of benchmark flows. To ensure some degree of generalizability to unseen flows, the relationships between various closure coefficients are orchestrated to yield reasonable fixed-point and bifurcation behavior in different asymptotic limits of turbulence. Thus, when some of the closure coefficients are unilaterally altered, the closure system of equations may be adversely affected. In ML-RANS, some of the closure coefficients are represented by ML functionals trained with data from chosen flows. 
This study addresses three features necessary for seamlessly blending traditional and ML elements in the closure framework:
\begin{enumerate}
\item {\bf Compatibility conditions}:  It is shown that the discordance between the ML-based constitutive closure coefficients (CCC) and traditional transport-equation closure coefficients (TCC) can lead to erroneous predictions. It is shown that this shortcoming can be overcome by imposing physics-based constraints among the various coefficients. 
\item{\bf Training consistency}: The inherent limitations of currently popular open-loop training are investigated. Specifically the lack of consistency between the baseline model closure coefficients that produce the input features and ML functional for the same coefficients is examined. It is demonstrated that an iterative closed-loop training procedure can lead to consistency between the equations that produce the input features and the output which is the ML functional. 
\item{\bf Loss function formulation}: The importance of formulating the most appropriate objective (or loss) function is examined. It is proposed that the loss function based on a combination of anisotropy tensor ($b_{ij}$) and Reynolds stress tensor ($\langle u_1u_2\rangle$) may be required to optimize the training outcome.
\end{enumerate}
The proposed compatibility condition, consistency procedure and loss function formulation are investigated in a simple channel flow. In the evaluation process, the standard $k-\omega$ model is intentionally altered from its original (correct) form and the ability of open and closed loop training methods to recover the original level of performance is examined. It is shown that the closed loop training process with compatibility constraints leads to significantly improved predictions over open loop method. Future work will focus on predictive computations of more complex flows and formulation of other compatibility constraints.

\clearpage 
\section*{Acknowledgement}
The authors acknowledge support provided by Texas A\&M High Performance Research Computing center.
\section*{References}
\bibliographystyle{ieeetr}
\bibliography{main}


\end{document}